\documentclass[journal=jacsat,manuscript=article, layout=twocolumn]{achemso}

\usepackage{chemformula} 
\usepackage[T1]{fontenc} 

\author{Kostas Kanellopulos}
\author{Robert G. West}
\author{Stefan Emminger}
\author{Paolo Martini}
\affiliation[TU Wien]
{Institute of Sensor and Actuator Systems, TU Wien, 1040 Vienna, Austria.}
\author{Markus Sauer}
\author{Annette Foelske}
\affiliation[TU Wien]
{Analytical Instrumentation Center, TU Wien, 1060 Vienna, Austria}
\author{Silvan Schmid}
\affiliation[TU Wien]
{Institute of Sensor and Actuator Systems, TU Wien, 1040 Vienna, Austria.}
\email{silvan.schmid@tuwien.ac.at}

\title{Stress-Dependent Optical Extinction in LPCVD Silicon Nitride Measured by Nanomechanical Photothermal Sensing}

\abbreviations{VIS,IR,NPAS}
\keywords{American Chemical Society, \LaTeX}

\begin{document}


\begin{abstract}
  Understanding optical absorption in silicon nitride is crucial for cutting-edge technologies like photonic integrated circuits, nanomechanical photothermal infrared sensing and spectroscopy, and cavity optomechanics. Yet, the origin of its strong dependence on film deposition and fabrication process is not fully understood. This Letter leverages nanomechanical photothermal sensing to investigate optical extinction $\kappa_{\mathrm{ext}}$ at 632.8 nm wavelength in LPCVD SiN strings across a wide range of deposition-related tensile stresses ($200-850$ MPa). Measurements reveal a reduction in $\kappa_{\mathrm{ext}}$ from 10$^3$ to 10$^1$ ppm with increasing stress, correlated to variations in Si/N content ratio. Within the band-fluctuations framework, this trend indicates an increase of the energy bandgap with the stress, ultimately reducing absorption. Overall, this study showcases the power and simplicity of nanomechanical photothermal sensing for low absorption measurements, offering a sensitive, scattering-free platform for material analysis in nanophotonics and nanomechanics.
\end{abstract}

Investigating the optical properties of solid-state materials is essential for both fundamental and applied science. Optical absorption, in particular, is critical in various fields, including photonic integrated circuits (PIC) for quantum information,\cite{Moody2022} and the design of nanomechanical resonant sensors for infrared (IR) light detection, \cite{Piller2023} photothermal spectromicroscopy, \cite{Kirchhof2023, Kanellopulos2023, West2023} and cavity optomechanics \cite{Wilson2012, Reinhardt2016, Gaetner2018, Mason2019, Huang2024}.

In IR sensing, high optical absorption is desired to enhance the sensor's specific detectivity. \cite{Snell2022, Piller2023} Conversely, applications like PICs, cavity optomechanics, and nanomechanical photothermal spectromicroscopy require minimal absorption to realize high-confinement waveguides,\cite{Corato2024} to prevent mechanical instability\cite{Metzger2008, Huang2024} and cavity bistability,\cite{An1997} and to mitigate photothermal back-action frequency noise introduced in the resonator \cite{kanellopulos2024comparative}, respectively.

Silicon nitride (SiN) holds a prominent position in these fields for its excellent mechanical, thermal, and optical properties.\cite{Ji2023, Kaloyeros2020} Its extensive use in photonics stems from its broad transparency window ($0.4-8\ \mathrm{\mu m}$), which, however, strongly depends on the film deposition and fabrication process, for which the underlying mechanisms are not fully understood. 
This has been observed by means of various characterization techniques, such as ellipsometry \cite{Makino1983, Poenar1997, Lin2013, Krueckel2017, Resende2021, Beliaev2022}, direct single-pass absorption spectroscopy \cite{Philipp_1973}, FTIR interferometry \cite{Cataldo2012, Yang2018}, cutback \cite{Sorace2019}, outscattered light method \cite{Smith2023, Blasco2024, Lelit2022, Mashayekh2021, Gorin2008}, prism coupling \cite{Bonneville2021}, photoluminescence \cite{Museur2016}, photothermal common-path interferometry \cite{Steinlechner2017}, and cavity-enhanced absorption spectroscopy \cite{Stambaugh2014, Ji2023, Corato2024}. 
However, these approaches often suffer from scattering losses and slow measurement time, which obscure the true absorption of SiN and make analyses prone to parasitic heating of the surroundings. 

Within this context, nanomechanical photothermal spectroscopy offers a robust solution to these challenges.\cite{Yamada2013, Bose2014, Andersen2016, Kurek2017, Kanellopulos2023, Luhmann2023, Land2024}
Here, a nanomechanical resonator detects directly absorption via resonance frequency shifts due to photothermal heating, insensitive to scattering. Upon illumination, the resultant temperature rise makes the initial tensile stress relax, leading to frequency detuning. Due to its high power sensitivity, fast thermal response, and versatility in sensor design, this technique has significantly advanced the characterization of low-loss materials \cite{West2023}.
 
In this Letter, nanomechanical photothermal sensing is employed to elucidate the relationship between absorption and residual tensile stress in low-pressure chemical vapour deposition (LPCVD) deposited SiN thin films. The extinction coefficient at 632.8 nm wavelength is measured from low stress ($\approx 200$ MPa), relevant to photothermal-based applications,\cite{West2023} to high stress ($> 800 $ MPa), relevant to cavity optomechanics \cite{Sementilli2022, Engelsen2024} and PIC design.\cite{Corato2024}
The thin films are patterned in a string geometry, which ensures high photothermal responsivity and fast response, as previously demonstrated.\cite{kanellopulos2024comparative} The experimental results reveal a reduction in extinction (from 10$^3$ to 10$^1$ ppm) with increasingly higher tensile stress, consistent with previously observed trends.\cite{Beliaev2022} The measurements are analyzed within the framework of the band-fluctuations model,\cite{Guerra2019} attributing the observed reduction to a blue-shift in the energy bandgap caused by a decrease in silicon-to-nitrogen (Si/N) content ratio. Overall, this study underscores the power and simplicity of nanomechanical photothermal sensing for the characterization of low-loss materials in nanophotonics and nanomechanics.

\begin{figure}
\includegraphics[width = 0.5\textwidth]{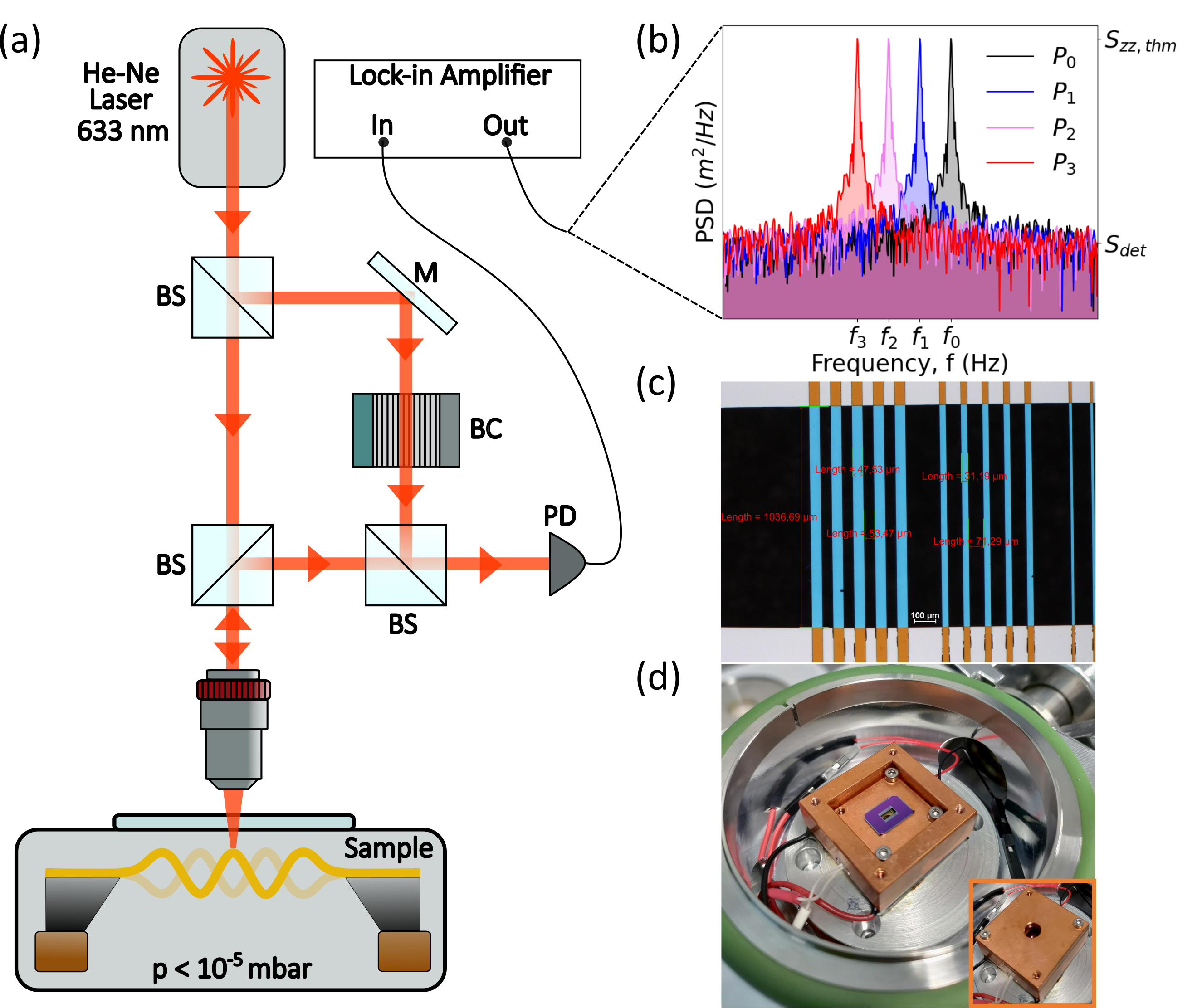}
\caption{(a) Sketch of the experimental set-up (LDV, Polytech GmbH MSA-500). A laser of wavelength $\lambda$ and input power $P_0$ impinges on the resonator, of absorption coefficient $\alpha_\mathrm{abs}(\lambda)$. This causes a frequency detuning of the nanomechanical resonator. BS: beam-splitter. BC: Bragg cell. PD: photodetector. (b) Mechanical frequency detuning measured by monitoring the shift of the thermomechanical noise peak of the string's fundamental mode as a function of $P_0$. (c) Optical micrograph of the SiN strings used in the present study. Orange/light blue regions are made of SiN; the grey regions are the Si substrate. (d) Photo of the cm-scale copper thermal equilibrium chamber used for the characterization of the linear coefficient of thermal expansion. A thermoelectric module is glued beneath to heat up the whole oven (thick red electrical connections) to guarantee a uniform temperature rise of the chips. The temperature is monitored and kept constant with a PID controller.}
\label{fig:setup}
\end{figure}

In the present setup, the nanomechanical resonator is operated in a custom-made vacuum chamber at high-vacuum conditions ($p<10^{-5}\ \mathrm{mbar}$) to reduce gas damping and thermal convection losses.\cite{Kanellopulos2023} The mechanical displacement is transduced optically via laser-Doppler vibrometry (LDV, Polytec GmbH MSA-500) equipped with a HeNe laser at 632.8 nm wavelength, with a beam waist of $\approx 1.5\ \mathrm{\mu m}$ (Fig.~\ref{fig:setup}). The same laser also probes SiN absorption, which simplifies the measurement procedure. For each structure, the frequency shift of the thermomechanical noise peak for the fundamental resonance mode is recorded at various optical powers ($6-120\ \mathrm{\mu W}$)\cite{Sadeghi2020, kanellopulos2024comparative}, as schematically shown in Fig.~\ref{fig:setup}b. A minimum of five resonators is evaluated for each stress and length. Fig.~\ref{fig:setup}c shows a representative image of the characterized strings.

The absorption coefficient $\alpha_\mathrm{abs}$ is determined via comparison between the theoretical $\mathcal{R}_{\mathrm{P}}$ and the experimental $\mathcal{R}_{\mathrm{P_0}}$ relative power responsivity for the string resonators.\cite{Schmid2023, West2023, kanellopulos2024comparative} 
On the one hand, $\mathcal{R}_{\mathrm{P}}$ denotes the relative frequency change per absorbed power $P(\lambda) = \alpha_{\mathrm{abs}}(\lambda) P_0$, and is expressed as \cite{Schmid2023}  
\begin{equation}\label{Rp_theory_general}
    \mathcal{R}_{\mathrm{P}}(\omega) = \frac{1}{f_0}\frac{\partial f_0}{\partial P} = \frac{\mathcal{R}_\mathrm{T}}{G}\left| H_\mathrm{th}(\omega)\right|,
\end{equation}
where $\mathcal{R}_\mathrm{T}$, $G$, and $H_\mathrm{th}(\omega)=(1 + \mathrm{i}\omega \tau_\mathrm{th})^{-1}$ denote the temperature responsivity in units [1/K], the thermal conductance in units [W/K], and the thermal response of the resonator, with $\tau_\mathrm{th}$ its thermal time constant, respectively.\cite{Schmid2023, kanellopulos2024comparative} All the measurements have been performed far from any thermal transient ($\omega\ll \tau_\mathrm{th}^{-1}$). Hence, the $\omega$-dependence is dropped in the following.

For a string resonator, Eq.~\eqref{Rp_theory_general} is given by \cite{Schmid2023}
\begin{equation}\label{Rp_theory}
    \mathcal{R}_{\mathrm{P}} = -\frac{\alpha_{\mathrm{th}}E}{2\sigma_0}\left[8 \frac{hw}{L} \kappa + 8Lw\epsilon_{\mathrm{rad}}\sigma_{\mathrm{SB}}T_0^3\right]^{-1},
\end{equation}
with $\alpha_{\mathrm{th}}$, $E$, $\sigma_0$, $h$, $w$, $L$, $\kappa$, $\epsilon_\mathrm{rad}$, $\sigma_\mathrm{SB}$ and $T_0$ denoting the resonator's linear coefficient of thermal expansion, Young's modulus, tensile stress, thickness, width, length, thermal conductivity, emissivity, Stefan-Boltzmann constant, and bath temperature, respectively. The thermal conductance $G$ includes the thermal dissipation through the surrounding frame $G_\mathrm{cond}$ (first addend in brackets), and thermal radiation to the environment $G_\mathrm{rad}$ (second addend in brackets).\cite{Schmid2023, kanellopulos2024comparative}

$\mathcal{R}_{\mathrm{P_0}}$ is obtained by directly measuring the relative frequency shift per impinging power $P_0$ (as schematically shown in Fig.~\ref{fig:setup}b). It relates to $\mathcal{R}_\mathrm{P}$ through the absorption coefficient $\alpha_{\mathrm{abs}}(\lambda)$ as follows
\begin{equation}\label{Rp_exp}
    \mathcal{R}_{\mathrm{P_0}}(\lambda) = \frac{1}{f_0} \frac{\partial f_0}{\partial P_0} = \frac{1}{f_0} \frac{\partial f_0}{\partial P} \frac{\partial P}{\partial P_0} = \alpha_\mathrm{abs}(\lambda)\ \mathcal{R}_\mathrm{P},
\end{equation}
with $\lambda$ denoting the probe optical wavelength. From Eq.~\eqref{Rp_exp}, it is possible to directly evaluate the optical absorption coefficient as $\alpha_\mathrm{abs}=\mathcal{R}_\mathrm{P_0}/\mathcal{R}_\mathrm{P}$.

\begin{figure*}
\includegraphics[width = 1\textwidth]{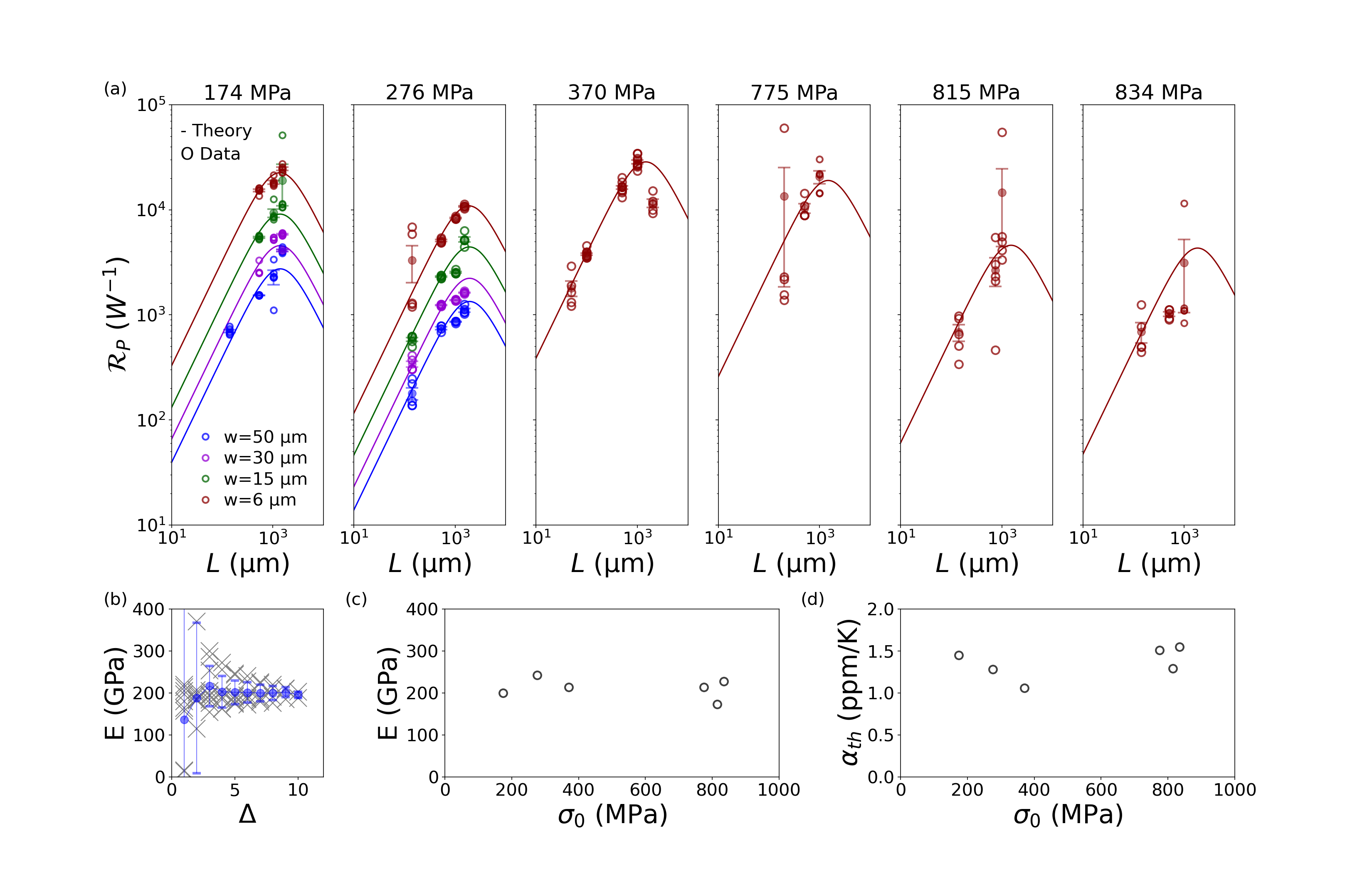}
\caption{(a) $\mathcal{R}_P$ for different SiN string structures. Circles: experimental responsivity \eqref{Rp_exp}, divided by the corresponding mean absorption coefficient $\alpha_\mathrm{abs}$. Solid curve: theoretical model \eqref{Rp_theory}. Material parameter assumed: $\rho = 3000\ \mathrm{kg/m^3}$, $\kappa=3$ W/(m K). Emissivity values are calculated from data reported in Ref.~\citenum{Cataldo2012}: 0.05 ($h=56$ nm), 0.13 ($h=157$ nm), 0.133 ($h=177$ nm), 0.171 ($h=312$ nm), 0.176 ($h=340$ nm). (b) Example of Young's modulus estimation, following the procedure of Ref.~\citenum{Klass2022}. (c) Experimental Young's modulus E as a function of the prestress $\sigma_0$. (d) Experimental linear coefficient of thermal expansion $\alpha_\mathrm{th}$ as a function of the prestress $\sigma_0$.}
\label{fig:Rp_sum}
\end{figure*}

The experimental power responsivity \eqref{Rp_exp} across the different stresses is displayed in Fig.~\ref{fig:Rp_sum}a as a function of the resonators' length $L$ (circles), together with the theoretical calculations \eqref{Rp_theory} (solid curves). The scale is given in terms of absorbed power $P$. It is worth noting that $\mathcal{R}_\mathrm{P}$ grows for longer strings in the conduction-limited regime ($L< 1$~mm), as $G\simeq G_\mathrm{cond}$ is inversely proportional to the length $L$ (see Eq.~\eqref{Rp_theory}), leading to better thermal insulation from the environment.\cite{kanellopulos2024comparative} Conversely, increasingly longer resonators ($L>1$ mm) enter the radiation-limited regime ($G\simeq G_\mathrm{rad}$), leading to a drop of $\mathcal{R}_\mathrm{P}$ as the radiating surface increases. 

\begin{table*}
    \caption{String resonators' geometrical ($h$), mechanical ($\sigma_0$, $E$, $\alpha_\mathrm{th}$), compositional (Si/N), and optical ($\eta$, $\kappa_\mathrm{ext}$, $E_\mathrm{g}$, $\beta^{-1}$) properties.}
    \label{tab:Strings}
    \begin{tabular}{|l|l|l|l|l|l|l|l|l|}
    \hline
    $\sigma_0$ ($\mathrm{MPa}$)&h ($\mathrm{nm}$)&E ($\mathrm{GPa}$)&$\alpha_{\mathrm{th}}$ ($\mathrm{ppm/K}$)& $\eta$ & $\kappa_\mathrm{ext}$ (ppm)& Si/N & $E_\mathrm{g}$ ($\mathrm{eV}$) & $\beta^{-1}$ ($\mathrm{meV}$)\\
    \hline
    174 & 177 & 200 & 1.45 & 1.215 & 606 & 0.96 & 3.23 & 201\\
    275 & 340 & 243 & 1.28 & 1.105 & 588 & 0.98 & 3.09 & 183\\
    370 & 56 & 214 & 1.06 & 1.022 & 176 & 0.89 & 3.62 & 212\\
    775 & 56 & 214 & 1.51 & 1.022 & 38 & 0.86 & 3.90 & 208\\
    815 & 157 & 173 & 1.29 & 1.273 & 20 & 0.83 & 4.21 & 227\\
    834 & 312 & 227 & 1.55 & 1.268 & 21 & 0.84 & 4.10 & 217\\
    \hline
    \end{tabular}
\end{table*}

The analyzed resonators have thicknesses of $h=56-340\ \mathrm{nm}$ and widths of $w=5-50\ \mathrm{\mu m}$, ensuring minimal thermal dissipation. As indicated in Eq.~\eqref{Rp_theory}, $G_\mathrm{cond}\propto h w$ and $G_\mathrm{rad}\propto w$, making these strings highly responsive to photothermal heating. Furthermore, the length $L$ varies in the range $0.1-2$ mm, making the resonator's power response mainly thermal conduction limited.\cite{kanellopulos2024comparative} The current experimental approach is, therefore, less influenced by the SiN emissivity, which, according to Kirchhoff's law, equals the optical absorption\cite{Bergman2017}—the parameter under scrutiny in this study. 
Moreover, the temperature responsivity $\mathcal{R}_\mathrm{T}$ appearing in Eq.~\eqref{Rp_theory} is independent of the Poisson’s ratio $\nu$, opposite to what occurs in, e.g., membrane resonators,\cite{Schmid2023, kanellopulos2024comparative} which further reduces the uncertainty in the absorption measurement stemming from its dependence on other material parameters.
Hence, the resonators employed here make this approach highly robust for solid-state material absorption characterization.

In this regard, the Young’s modulus $E$ and the linear coefficient of thermal expansion $\alpha_\mathrm{th}$ have been measured to reduce the uncertainty on the estimation of the absorption coefficient $\alpha_\mathrm{abs}$. $E$ has been estimated following the procedure of Ref.~\citenum{Klass2022}, upon recording of the strings' eigenmode spectrum (an example is displayed in Fig.~\ref{fig:Rp_sum}b, where $\Delta=|m - n|$, with $n \neq m$ being modal numbers). The experimental results are displayed in Fig.~\ref{fig:Rp_sum}c and Table~\ref{tab:Strings}, with values in the range $170-250$~GPa, consistent with previously reported data.\cite{Klass2022, Kaloyeros2020}

$\alpha_\mathrm{th}$ has been measured by recording the frequency shift of the thermomechanical noise peak as a function of controlled temperature rises ($\Delta T=0-10$ K),\cite{Larsen2011} through the relation $\alpha_\mathrm{th,SiN}=\alpha_\mathrm{th,Si} - 2 \mathcal{R}_\mathrm{T}\sigma_0/E$. For that, a thermoelectric module (GM200-127-10-15, Adaptive Power Management) has been used to heat up the resonators, while monitoring and keeping the temperature at the desired value via a PID controller (TEC-1092, Meerstetter Engineering). The chips have been enclosed inside a cm-scale copper thermal bath to guarantee thermal equilibrium through radiative heat transfer of the string with the environment (see Fig.~\ref{fig:setup}d). Fig.~\ref{fig:Rp_sum}d and Table~\ref{tab:Strings} show the corresponding results, with $\alpha_\mathrm{th}$ lying in the range $1-1.6\ \mathrm{ppm/K}$ (assuming $\alpha_\mathrm{th, Si}=2.6$ ppm/K), which are consistent with previously reported values.\cite{Larsen2011, Kaloyeros2020} No stress dependence has been observed for the two material parameters.

\begin{figure*}
\includegraphics[width = 1\textwidth]{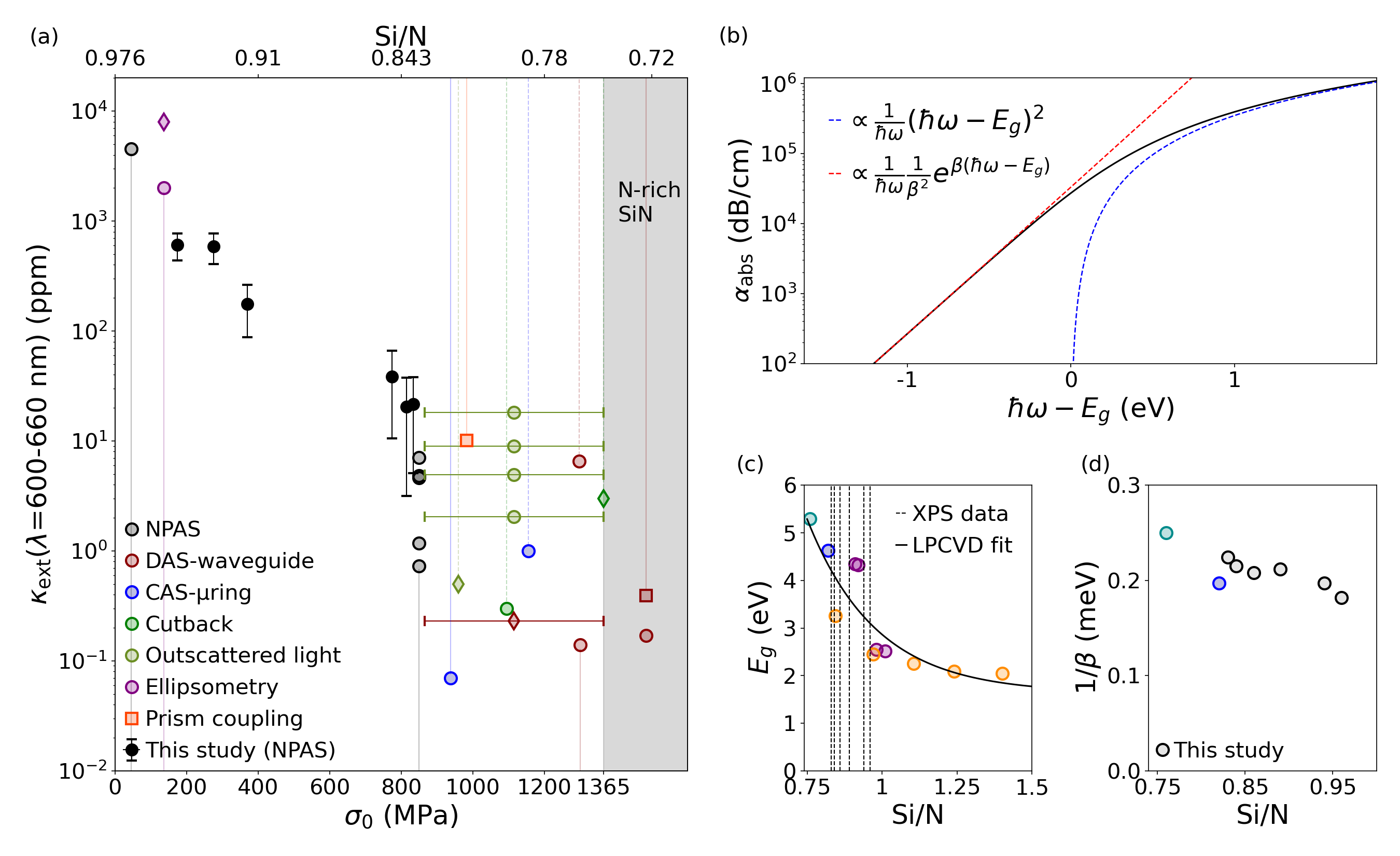}
\caption{(a) $\kappa_{ext}$ for different SiN string's tensile stresses at an excitation wavelength of $\lambda=(632.8 \pm 30)$ nm. Characterization techniques included in the figure are: nanomechanical photothermal absorption spectroscopy (NPAS) \cite{Land2024}, direct absorption spectroscopy (DAS) in waveguides \cite{Inukai1994, Bulla1999, Daldosso2004, Sacher2019}, cavity absorption spectroscopy in microrings resonators (CAS-µring) \cite{Corato2024, Worhoff2008}, cutback \cite{Sorace2019}, ellipsometry \cite{Poenar1997}, and prism coupling \cite{Bonneville2021}. Markers refer to LPCVD (circles), plasma-enhanced CVD (PECVD, diamonds), and electron-cyclotron resonance CVD (ECR-CVD, squares) deposited SiN films. For the reported values, the vertical lines indicate a relationship with stress $\sigma_0$ (intersection with the bottom x-axis) or Si/N (intersection with the top x-axis), explicitly given in (solid lines) or derived from (dashed lines) the original article. When none of these values could be extracted, a stress error bar has been used ($\sigma_0=865-1365$ MPa).
(b) Absorption coefficient in the band-fluctuations model. The dashed blue and red curves represent the absorption due to electronic transition between extended states (Tauc regime), and absorption due to disorder-induced localized to extended state transitions (Urbach regime), respectively. 
(c) Energy bandgap $E_\mathrm{g}$ as a function of the Si/N ratio. The solid curve is a fitting function of the displayed reported values of the form $f(x)=a e^{b x} + c$, with $a= 95.94$ eV, $b=4.356$, and $c=1.633$ eV. Only LPCDV SiN films have been considered. Compilation: darkcyan, Ref.~\citenum{Bauer1977}; blue, Ref.~\citenum{Corato2024}; purple, Ref.~\citenum{Beliaev2022}; orange, Ref.~\citenum{Krueckel2017}. Dashed vertical lines indicate the Si/N ratios measured in this study with XPS. Intersections with the fitting curve are given in Table~\ref{tab:Strings}. (d) Corresponding Urbach energy $\beta^{-1}$ of the thin films analyzed in this study (black circles). For comparison, data from Ref.~\citenum{Bauer1977} (darkcyan) and Ref.~\citenum{Corato2024} (blues) are displayed.}
\label{fig:kext}
\end{figure*}

From the absorption measurements, the extinction coefficient for these thin films can be evaluated as \cite{Macleod2010}
\begin{equation}\label{eq_kext}
    \kappa_\mathrm{ext}(\lambda) =\ \frac{\lambda\ \alpha_\mathrm{abs}(\lambda) }{4\pi\ h\ \eta}.
\end{equation}
with $\eta$ denoting a dimensionless factor that accounts for possible interference inside the thin SiN slab.\cite{Land2024} For the film thicknesses analyzed here, $\eta\approx 1 - 1.27$ at 632.8 nm wavelength (see Table~\ref{tab:Strings} and Supporting Information).
Fig.~\ref{fig:kext}a shows the nanomechanical photothermal results of $\kappa_\mathrm{ext}$ as a function of the resonators' tensile stress $\sigma_0$ (black circles). $\kappa_\mathrm{ext}$ decreases from $\approx 10^3$ ppm for the lowest stress, to $\approx 10^1$ ppm for the highest. These findings are compared with previously reported values of optical extinction for LPCVD (colored circles), as well as PECVD (colored diamonds), and ECR-CVD (colored squares) deposited SiN films (see Supporting Information for details on their deposition dependencies). The variance in magnitude among the compiled data for $\sigma_0\geq 850$~MPa can be partially attributed also to the inability of some of the considered techniques to differentiate between true absorption and scattering losses (in particular cutback and outscattered light \cite{Corato2024}).
Overall, a general trend emerges in Fig.~\ref{fig:kext}a, with $\kappa_\mathrm{ext}$ decreasing for increasingly higher SiN deposition-related tensile stress.

The measurements are analyzed within the framework of the band-fluctuations model,\cite{Guerra2019} which describes the absorption coefficient in units of [dB/m] as a function of the excitation energy $\hbar \omega$ for amorphous materials as
\begin{equation}\label{eq_abs_model}
    \alpha_\mathrm{abs}(\hbar \omega) =\ \frac{\alpha_0}{\hbar \omega} \frac{1}{\beta^2} \mathcal{J}_\mathrm{cv}(\beta(\hbar \omega - E_\mathrm{g})),
\end{equation}
where $\alpha_0$, $\beta$, $E_\mathrm{g}$, and $\mathcal{J}_\mathrm{cv}$ denote a coefficient collecting physical constants in unit [$\mathrm{(m\ eV)^{-1}}$], the Urbach slope in units [$\mathrm{eV}^{-1}$], the energy bandgap in units [eV], and a dimensionless joint electronic density of states (DOS), respectively. Fig.~\ref{fig:kext}b displays its functional form. For excitation energies $\hbar\omega>E_\mathrm{g}$, Eq.~\eqref{eq_abs_model} converges to the Tauc regime \cite{Beliaev2022}, where only fundamental electronic transitions between extended states are considered (dashed blue curve); for $\hbar \omega<E_\mathrm{g}$, the model converges to the empirical Urbach tail, where electronic defect-induced absorption follows $\alpha_\mathrm{abs} \propto e^{\beta \hbar \omega}$ (dashed red curve) \cite{Guerra2019}.

The model input parameters $E_\mathrm{g}$ and the Urbach energy $\beta^{-1}$ depend on the film deposition process through the residual tensile stress present in the films. In turn, this dependence is underpinned by the underlying correlation between the stress and the corresponding Si/N ratio, with the former increasing as the latter is reduced (see Table~\ref{tab:Strings}), as observed in LPCVD, as well as PECVD and ECR-CVD deposited SiN films \cite{Temple1998, Krueckel2017, Beliaev2022}.
Hence, the optical extinction reduction observed in Fig.~\ref{fig:kext}a for increasing tensile stress has to be related to the difference in the chemical composition of the thin films.

In this regard, the Si/N ratio of each chip has been experimentally characterized by X-ray photoelectron spectroscopy (XPS, PHI Versa Probe III-spectrometer) equipped with a monochromatic Al-$\mathrm{K_{\alpha}}$ X-ray source and a hemispherical analyser. Data analysis was performed using CASA XPS and Multipak software packages. The results are displayed in Table~\ref{tab:Strings}, and are consistent with those reported in previous works for similar tensile stress range.\cite{Beliaev2022, Temple1998, Yang2018} These values are also shown in the top x-axis of Fig.~\ref{fig:kext}a, to highlight how SiN extinction increases with Si/N.

Finally, the energy bandgap $E_\mathrm{g}$ of each thin film has been extracted by means of the fitting curve constructed from the compilation of previous works on LPCVD SiN only,\cite{Bauer1977, Corato2024, Krueckel2017, Beliaev2022} which are shown in Fig.~\ref{fig:kext}c. The XPS data (dashed vertical lines) are shown for clarity, and fall in the region of strongest dependence on Si/N. The corresponding energy bandgap (Table~\ref{tab:Strings}) has been found to increase from $\approx 3$~eV,  for the highest relative Si concentration, to $\approx 4.2$~eV, for the lowest. All these values exceed the probing energy used in this study ($\hbar \omega = 1.96\ \mathrm{eV}$), indicating that the absorption results from localized-to-extended electronic transitions of disorder-induced tail states, as it occurs typically in amorphous semiconductors.\cite{Guerra2019} 

With the energy bandgap defined for each thin film, the corresponding Urbach energy $\beta^{-1}$ has been determined by matching the experimental absorption to the band-fluctuations model. The results are shown in Fig.~\ref{fig:kext}d and Table~\ref{tab:Strings}, and are consistent with previously reported studies of LPCVD SiN ($\beta^{-1}\approx 200$ meV).\cite{Bauer1977, Corato2024} $\beta^{-1}$ slightly decreases with increasing Si/N ratios, as it has been observed also for PECVD deposited SiN, but at lower values (see the Supporting Information for a comparison) \cite{Garcia1995, Kato2003}.
Hence, lowering the Si/N ratio has the main effect of shifting the bandgap $E_g$ to higher energies, broadening the SiN transparency window. Conversely, the Urbach energy $\beta^{-1}$ does not vary significantly among these thin films, indicating that the reduction in extinction coefficient $\kappa_\mathrm{ext}$ is driven by an exponential decrease in the disorder-induced electronic tail DOS at the probing energy of 1.96~eV.

In conclusion, it has been shown that nanomechanical photothermal spectroscopy represents a highly sensitive, simple, and scattering-free platform for the optical characterization of low-loss materials. In this study, its capabilities have been explored using nanostring resonators made of LPCVD deposited SiN. Upon meticulous characterization of their mechanical and thermomechanical properties, it has been shown that SiN intrinsic extinction coefficient decreases with increasingly higher thin film tensile stress. This trend is attributed to a blue-shift in energy bandgap as a function of material composition. Therefore, varying the Si/N ratio provides a degree of freedom to tune the optical properties of SiN, advancing the understanding of this ubiquitous material.

\begin{suppinfo}
The Supporting information is available free of charge.
\begin{itemize}
  \item Calculation of the emissivity and factor $\eta$ for different thicknesses; compilation of data (table); derivation of Si/N ratios for the compiled data; comparison of the measured Urbach energies with reported data for LPCVD and PECVD SiN films.
\end{itemize}
\end{suppinfo}

\begin{acknowledgement}
The authors thank Johannes Hiesberger for the help with the experimental set-up, and Hajrudin Besic, Nicola Cavalleri and Niklas Luhmann for useful discussions. 
This project received funding from the Novo Nordisk Foundation under project MASMONADE with project number NNF22OC0077964.
Moreover, the Austrian Research Promotion Agency (FFG) is gratefully acknowledged for funding of the used XPS infrastructure (FFG project number: 884672).
\end{acknowledgement}

\bibliography{sin_abs_biblio}

\end{document}


\section{Emissivity $\epsilon_\mathrm{rad}$}
\begin{figure}
\includegraphics[width = 0.5\textwidth]{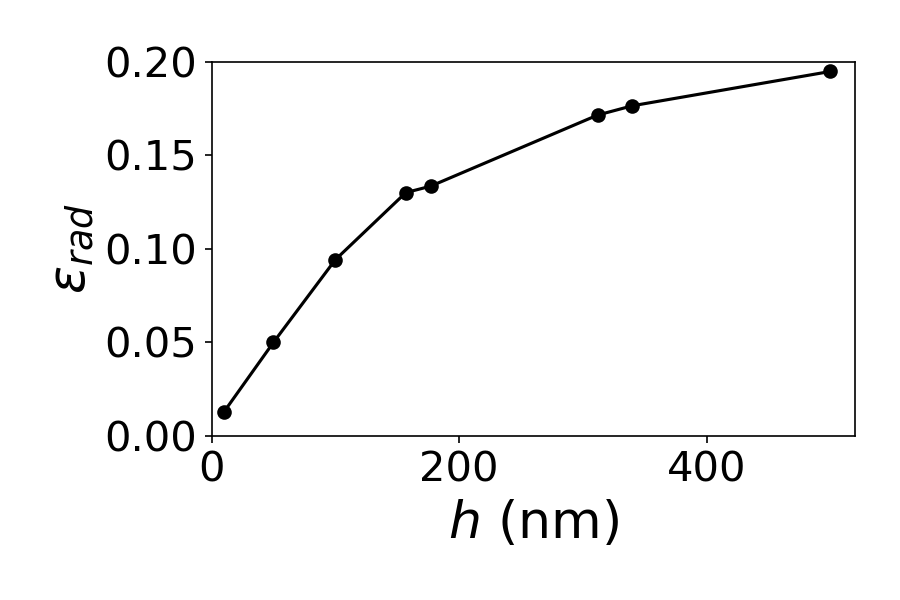}
\caption{Emissivity $\epsilon_\mathrm{rad}$ as a function of the film thickness $h$.}
\label{fig:emissivity}
\end{figure}

The values of emissivity used for $\mathcal{R}_\mathrm{P}$ in the main text are based on the experimental data of Ref.~\citenum{Cataldo2012}. The calculations are based on the matrix formalism for a single film surrounded by vacuum \cite{Macleod2010}, equivalent to what done in Ref.~\citenum{Zhang2020}. The results are shown in Fig.~\ref{fig:emissivity}, with $\epsilon_\mathrm{rad}$ increasing with the thickness, as expected for dielectric materials \cite{Edalatpour2013}.

\section{Calculations of $\eta$}
\begin{figure}
\includegraphics[width = 0.5\textwidth]{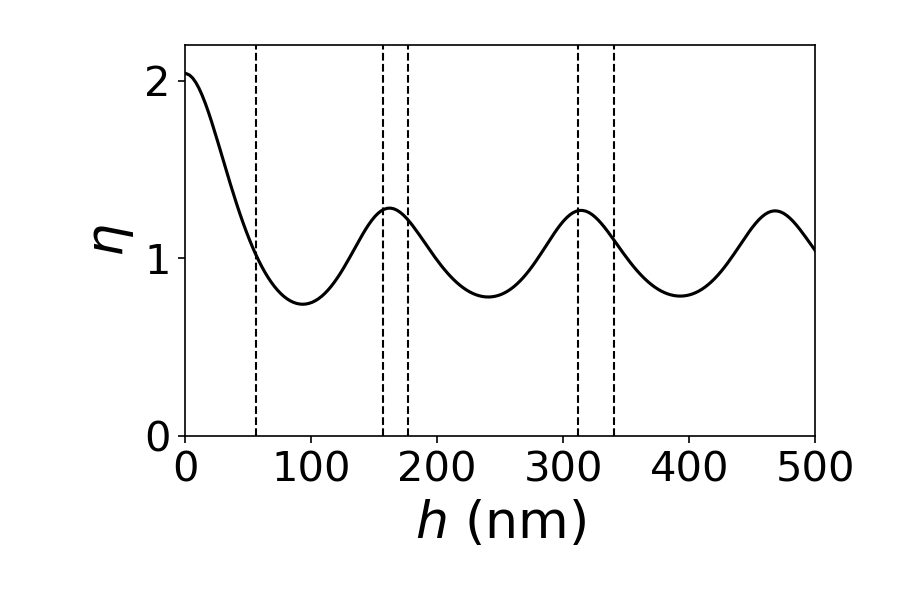}
\caption{Factor $\eta$ as a function of the film thickness $h$ at 632.8 nm wavelength. The solid curve is calculated with Eq.~\eqref{eq_eta}. The dashed vertical lines indicate the thicknesses of the films analyzed in this study.}
\label{fig:eta}
\end{figure}

The factor $\eta$ appearing in Eq.~(4) in the main text accounts for the interference effect inside the thin film. The calculations for the values presented in the Table~1 are based on Eq.~8 of Ref.~\citenum{Bubenzer1984}, equivalent to what shown in Ref.~\citenum{Land2024}. For a thin film of thickness $h$ and complex refractive index $\Tilde{n}=n + i \kappa_{ext}$, probed at a wavelength $\lambda$, the absorption correction factor is given by \cite{Land2024}
\begin{equation}\label{eq_eta}
    \eta = \frac{4n(n^2+1)+(n^2-1)\frac{\lambda}{\pi h} sin(\frac{4\pi n h}{\lambda})}{1+6n^2+n^4-(n^2 - 1)^2 cos(\frac{4\pi n h}{\lambda})}.
\end{equation}
The dispersive part of the refractive index, $n$, is obtained from the Sellmeier equation \cite{Luke2015}
\begin{equation}\label{eq_nSellmeier}
    n(\lambda)^2 = 1 + \frac{3.0249 \lambda^2}{\lambda^2 - 135.3406^2} + \frac{40314 \lambda^2}{\lambda^2 - 1239842^2},
\end{equation}
with $\lambda$ given in units of nanometer. For $\lambda = 632.8$ nm, $n=2.04$. Fig.~\ref{fig:eta} shows how $\eta$ varies with the film thickenss $h$ at this wavelength. The values corresponding to the thicknesses used in this work are summarized in Table~1 in the main text.

\section{Data compilation}
\begin{table*}
    \caption{List of the extinction values and corresponding parameters used in the main text ($\lambda = 600 - 660$ nm).}
    \label{tab:compilation}
    \begin{tabular}{|l|l|l|l|l|l|l|l|}
    \hline
    Reference&Deposition&Method&$\lambda$ (nm)& $\kappa_\mathrm{ext}$ (ppm)& $\sigma_0$ (MPa) & Si/N \\
    \hline
    \citenum{Land2024} & LPCVD & NPAS & 632.8 & 0.73 & 850 & -\\
     & & & & 1.17 & & \\
     & & & & 4.64 & & \\
     & & & & 7.07 & & \\
     & & & & 4.84 & & \\
    \hline
    \citenum{Inukai1994} & LPCVD & DAS-waveguide & 630 & 0.17 & - & 0.73 \\
     & ECR-CVD & & & 0.39 &  & \\
    \hline
    \citenum{Bulla1999} & PECVD & DAS-waveguide & 632.8 & 0.23 & - & - \\
    \hline
    \citenum{Corato2024} & LPCVD & CAS-µring & 644 & 0.07 & - & 0.82\\
    \hline
    \citenum{Worhoff2008} & LPCVD & CAS-µring & 632.8 & 1 & - & 0.783\textsuperscript{\emph{a}}\\
    \hline
    \citenum{Sorace2019} & LPCVD & Cutback & 643 & 0.3 & - & 0.794\textsuperscript{\emph{a}}\\
     & PECVD & & & 3 & & 0.751\textsuperscript{\emph{b}}\\ 
    \hline
    \citenum{Blasco2024} & LPCVD & Outscattered light & 633 & 18.134 & - & - \\
    \hline
    \citenum{Lelit2022} & LPCVD & Outscattered light & 660 & 8.98 & - & - \\
    \hline
    \citenum{Smith2023} & LPCVD & Outscattered light & 630 & 4.91 & - & - \\
    \hline
    \citenum{Mashayekh2021} & LPCVD & Outscattered light & 640 & 2.037 & - & - \\
    \hline
    \citenum{Gorin2008} & PECVD & Outscattered light & 633 & 0.5& - & 0.817\textsuperscript{\emph{b}}\\
    \hline
    \citenum{Sacher2019} & LPCVD & DAS-waveguide & 648 & 6.54 & - & 0.761\textsuperscript{\emph{a}} \\
    \hline
    \citenum{Poenar1997} & LPCVD & Ellipsometry & 633 & 2000 & 137.5 & - \\
     & PECVD & & & 8000 & & \\
    \hline
    \citenum{Bonneville2021} & ECR-CVD & Prism coupling & 632.8 & 10.15 & - & 0.813\\
    \hline
    \end{tabular}
    
    \textsuperscript{\emph{a}} Value derived from the dispersive refractive index $n_\mathrm{LPCVD}$ for LPCVD.\\
    \textsuperscript{\emph{b}} Value derived from the dispersive refractive index $n_\mathrm{PECVD}$ for PECVD.
\end{table*}
For some of the articles listed in Table~\ref{tab:compilation}, the derivation of the values is based on the existing relationship between the dispersive part of refractive index $n$ and the Si/N ratio, as shown already \cite{Makino1983, Maeckel2002}. The derivation is developed for each reference, where needed.

\subsection{Derived parameters for LPCVD}
\begin{figure}
\includegraphics[width = 0.5\textwidth]{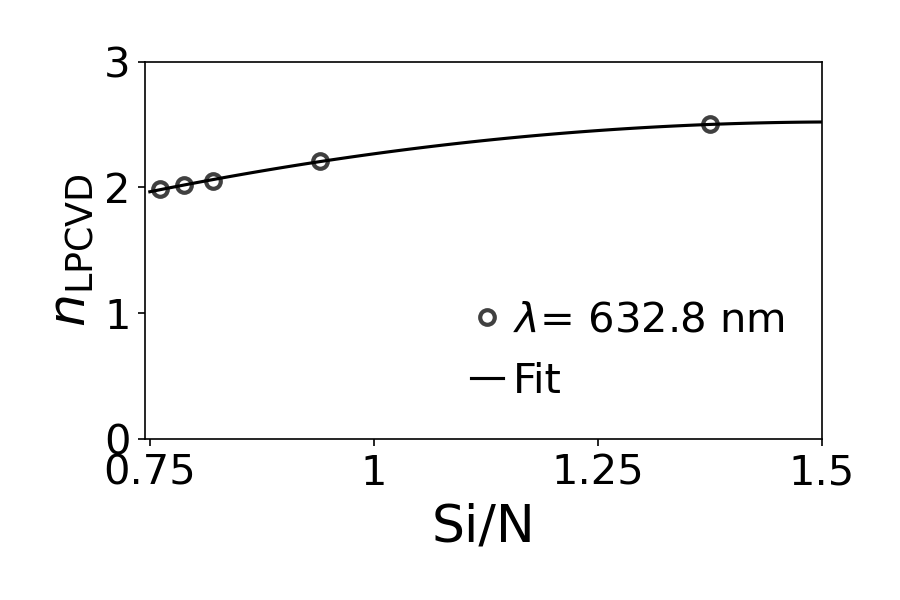}
\caption{Refractive index as a function of the Si/N ratio in LPCVD SiN, measured at 632.8 nm wavelength (circles). The solid curve is the fit used to derive the Si/N ratio.}
\label{fig:n_makino}
\end{figure}

For LPCVD SiN thin films, the measurements carried out in Ref.~\citenum{Makino1983} are used to derive the Si/N ratio from $n_\mathrm{LPCVD}$ at 632.8 nm wavelength. Fig.~\ref{fig:n_makino} shows the reported value (circles), together with the corresponding fit. The latter has the form $f(x) = p_1 x^2 + p_2 x + p_3$, with $p_1=-0.9333$, $p_2=2.839$, and $p_3=0.3608$.

\subsubsection{Reference~\citenum{Worhoff2008}}
It is reported a value of $n_\mathrm{LPCVD}=2.0115$ for TE mode at 632.8 nm wavelength.

\subsubsection{Reference~\citenum{Sorace2019}}
It is reported the following Sellmeier equation (upon fitting of the ellipsometric data)
\begin{equation}\label{n_lpcvd_Sorace}
    n_\mathrm{LPCVD} = \sqrt{1 + \frac{2.926 \lambda^2}{\lambda^2 - 23.47\cdot10^{-15}}}
\end{equation}
Hence, $n_\mathrm{LPCVD}(643\ \mathrm{nm})=2.02536$.

\subsubsection{Reference~\citenum{Sacher2019}}
It is reported a value of $n_\mathrm{LPCVD}(648\ \mathrm{nm)} = 1.98$.

\subsection{Derived parameters for PECVD}
For PECVD SiN thin film, the formula from Ref.~\citenum{Maeckel2002} has been exploited. In particular, the following relations holds
\begin{equation}\label{n_pecvd_Si_N}
    \mathrm{\frac{Si}{N}} = \frac{3}{4} \frac{n_\mathrm{PECVD} + n_\mathrm{a-Si:H} - 2 n_\mathrm{a-Si_3 N_4}}{n_\mathrm{a-Si:H} - n_\mathrm{PECVD}},
\end{equation}
with $n_\mathrm{PECVD}$, $n_\mathrm{a-Si:H}(632.8\ \mathrm{nm})=3.3$, and $n_\mathrm{a-Si_3 N_4}(632.8\ \mathrm{nm})=1.9$ denoting the measured refractive index, the index for a-Si:H, and for the stoichiometric SiN, respectively \cite{Maeckel2002}.

\subsubsection{Reference~\citenum{Sorace2019}}
The Sellmeier equation obtain by fitting the ellipsometry results has been given as
\begin{equation}\label{n_pecvd_Sorace}
    n_\mathrm{PECVD} = \sqrt{1 + \frac{2.503\lambda^2}{\lambda^2 - 17.29\cdot10^{-15}}}
\end{equation}
Hence, $n_\mathrm{PECVD}(643\ \mathrm{nm})=1.9006$.

\subsubsection{Reference~\citenum{Gorin2008}}
It is reported a value of $n_\mathrm{PECVD}(633\ \mathrm{nm})=1.96$.

\section{SiN Urbach energy: LPCVD vs PECVD}
\begin{figure}
\includegraphics[width = 0.5\textwidth]{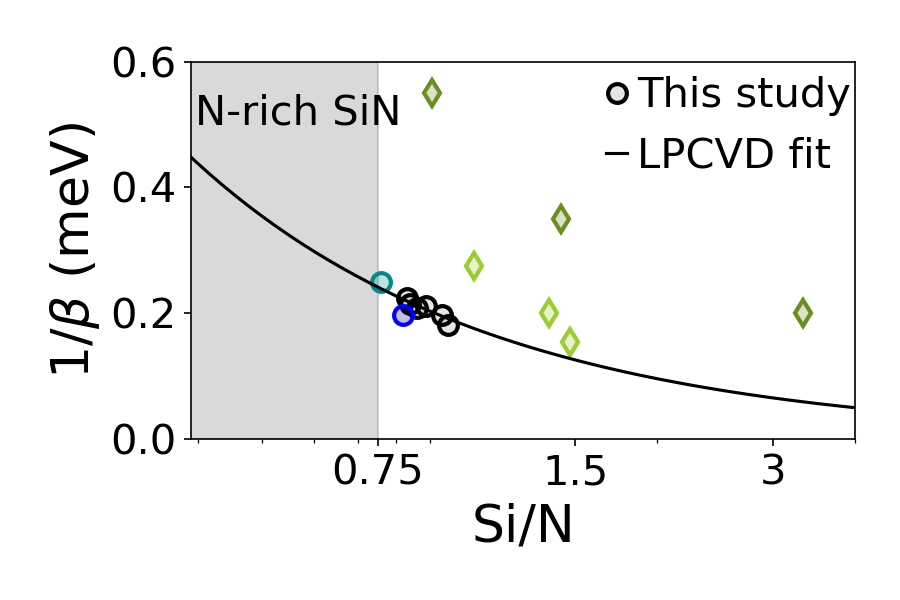}
\caption{Urbach energy $\beta^{-1}$ as a function of the Si/N ratio, for LPCVD (circles) and PECVD (diamond) SiN films. The compilation of data includes: blue, Ref.\citenum{Corato2024}; darkcyan, Ref.~\citenum{Bauer1977}; dark green, Ref.~\citenum{Garcia1995}; light green, Ref.~\citenum{Kato2003}. The solid curve is a fitting curve of the form $f(x) = a e^{b}$ for the LPCVD data only, with $a = 0.1843$ meV and $b=-0.9427$.}
\label{fig:beta}
\end{figure}

The Urbach energies $\beta^{-1}$ discussed in the main text are for LPCVD SiN only. Fig.~\ref{fig:beta} shows a comparison with PECVD films (diamond). These films present higher values of $\beta^{-1}$ than LPCVD ones for similar Si/N values. This means that absorption due to electronic transitions between disorder-induced localized to extended states is increased in the former. Overall, $\beta^{-1}$ reduces with Si/N for both deposition methods.

\bibliography{sin_abs_biblio}